\documentclass[a4paper]{easychair}

\SetMathAlphabet{\mathcal}{normal}{OMS}{xmdcmsy}{m}{n}
\hypersetup{%
  colorlinks=true,%
  citecolor=blue,%
  urlcolor=blue,%
  linkcolor=blue,%
}

\usepackage{isabelle}
\usepackage{stmaryrd}
\usepackage{tikz}
\usepackage{amssymb}
\usepackage{xspace}
\usepackage{ifthen}
\usepackage{amsmath}  
\usepackage{array} 
\usepackage[english]{babel}
\usepackage{cite}

\newtheorem{theorem}{Theorem}

\newtheorem{definition}[theorem]{Definition}
\newtheorem{example}[theorem]{Example}

\renewcommand{\theenumi}{\roman{enumi}}

\makeatletter
\renewcommand{\p@enumii}{\theenumi--}
\makeatother

\newcommand{\rSub}[1]{(\ref{#1})}

\newcommand\TTTT{%
 \ensuremath{\textsf{T\kern-0.2em\raisebox{-0.3em}T\kern-0.2emT\kern-0.2em\raisebox{-0.3em}2}}%
}
\newcommand{\aprove}{{\sf APro\kern-0.1emVE}}

\newcommand\ceta{\textsf{C\kern-0.2exe\kern-0.5exT\kern-0.5exA}\xspace}
\newcommand\isafor{\textsf{Isa\kern-0.2exF\kern-0.2exo\kern-0.2exR}\xspace}
\newcommand\cocci{{\sf Coccinelle}}
\newcommand\CoLoR{{\sf CoLoR}}
\newcommand\cime{{\sf CiME}}
\newcommand\rainbow{{\sf Rainbow}}
\newcommand\arcnats{\mathbb{A}_{\nats}}
\newcommand\arcints{\mathbb{A}_{\ints}}
\newcommand\arcrats{\mathbb{A}_{\rats}}
\newcommand\arczero{-\infty}
\newcommand\arcone{0}
\newcommand\nats{\mathbb{N}}
\newcommand\rats{\mathbb{Q}}
\newcommand\ints{\mathbb{Z}}

\newcommand\FF{\mathcal{F}}

\newcommand\RR{\mathcal{R}}
\newcommand\UU{\mathcal{U}}
\newcommand\VV{\mathcal{V}}

\newcommand\Pol{\mathcal{P}ol}

\newcommand\SN[1][]{\mathsf{SN}_{#1}}

\newcommand\rstep[1][\RR]{\to_{#1}}

\newcommand{\matm}[4]{\left(\begin{array}{cc}#1 & #2 \\ #3 & #4\end{array}\right)}

\newcommand\Ff{\mathsf{f}}

\newcommand\Fhalf{\mathsf{half}}

\newcommand\Fp{\mathsf{p}}

\newcommand\Fs{\mathsf{s}}

\newcommand\defref[2][]{Def.~\ref{#2}\ifthenelse{\equal{#1}{}}{}{\parref{#1}}}
\newcommand\exref[1]{Ex.~\ref{#1}}
\newcommand\lemref[2][]{Lem.~\ref{#2}\ifthenelse{\equal{#1}{}}{}{\parref{#1}}}

\newcommand\parref[1]{{(\ref{#1})}}

\newcommand\secref[1]{Sec.~\ref{#1}}

\newcommand\thmref[2][]{Thm.~\ref{#2}\ifthenelse{\equal{#1}{}}{}{\parref{#1}}}

\let\OLDparagraph\paragraph
\renewcommand\paragraph[1]{\OLDparagraph{\sl#1\/}}

\renewcommand{\theenumi}{\textsf{\alph{enumi}}}

\newcommand{\resid}[1]{\href{http://termcomp.uibk.ac.at/termcomp/competition/resultDetail.seam?resultId=#1}{#1}}

\begin{document}

\title{Certification extends Termination Techniques}
\author{
Christian Sternagel\thanks{%
  This author is supported by FWF (Austrian Science Fund) project P18763.
}~~(\url{christian.sternagel@uibk.ac.at}), University of Innsbruck, Austria \\
\and Ren\'e Thiemann (\url{rene.thiemann@uibk.ac.at}), University of Innsbruck, Austria 
}
\authorrunning{
Sternagel, Thiemann}

\maketitle




\section{Introduction}

Termination provers for term rewrite systems (TRSs) became more and more
powerful in the last years.  One reason is that a proof of termination no
longer is just some reduction order which contains the rewrite relation of
the TRS.  Currently, most provers combine basic termination
techniques in a flexible way using the dependency pair
framework (DP framework) or rule removal.  Hence, a termination proof is a tree where
at each node a specific technique is applied. Therefore, instead of just
stating the precedence of some lexicographic path order or giving some
polynomial interpretation, current termination provers return proof trees
consisting of many different techniques and reaching sizes of several
megabytes. Thus, it would be too much work to check by hand whether these
trees really form a valid proof. (Also, checking by hand does not provide a
very high degree of confidence.)

It is regularly demonstrated that we cannot blindly trust in the output of
termination provers.  Every now and then, some termination prover delivers
a faulty proof. Most often, this is only detected if there is another
prover giving a contradicting answer on the same problem.  To improve this
situation, three systems have been developed over the last few years:
\cime/\cocci{} \cite{A3PAT2,A3PAT}, \rainbow/\CoLoR{} \cite{CoLoR}, and
\ceta/\isafor{} \cite{CeTA}. These systems either certify or reject a given
termination proof.  Here, \cocci{} and \CoLoR{} are libraries on rewriting
for Coq (\url{http://coq.inria.fr}) and \isafor{} is our library on
rewriting for Isabelle~\cite{Isabelle}. (Throughout this paper we just
write Isabelle whenever we refer to Isabelle/HOL.) And indeed, using
certifiers several bugs have been detected. For example, in the
termination competition of the last year (November 2009), at least eight 
faulty proofs were spotted
by certifiers.\footnote{Result
IDs 
\url{http://termcomp.uibk.ac.at/termcomp/competition/resultDetail.seam?resultId=135160}, 
\resid{136252}, 
\resid{136278}, 
\resid{136365}, 
\resid{136378},
\resid{136499}, 
\resid{137163}, and
\resid{137465}
}
(Caused by three different bugs, all of which were most likely due to some
output error.)

Although many termination techniques have already been formalized---\ceta
can certify termination or nontermination proofs for 1522 out of the 2132
TRSs from the TPDB version 7.0.2 which is over 70 \% of the whole
database---there are still several techniques that have not been
formalized. So, clearly there are termination proofs that are produced by
some termination tool where the certifiers have to become more powerful.

However, a similar situation also occurs in the other direction. We have formalized
termination techniques in a more general setting as they have been introduced.
Hence, currently we can certify proofs using techniques that 
no termination tool supports so far. In this paper we shortly present
two of these formalizations.
\begin{enumerate}
\item\label{np} Polynomial orders with negative constants \cite{HM07}.
\item\label{arc} Arctic termination \cite{Arctic}.
\end{enumerate}

Here, for \rSub{np} we were able to lift the result from the naturals as
introduced in \cite{HM07} to an arbitrary carrier, including matrices
(\secref{snp}).  For \rSub{arc} we have generalized the arctic semiring and
the arctic semiring below zero into one semiring which subsumes both
existing approaches and extends them to the rationals (\secref{sarc}).

Note that all the proofs that are presented (or omitted) in the following,
have been formalized in our Isabelle library \isafor.  This library and the
executable certifier \ceta{} are available at \ceta's website:
\begin{center} 
\url{http://cl-informatik.uibk.ac.at/software/ceta}
\end{center}

\section{Preliminaries}
\label{s:preliminaries}

We assume  familiarity with term rewriting \cite{BN98}. Still, we
recall the most important notions that are used later on. A 
\emph{term}~$t$ over a set of \emph{variables}~$\VV$ and a set of
\emph{function symbols}~$\FF$ is
either a variable $x \in \VV$ or an \mbox{$n$-ary} function symbol~$f \in \FF$ applied to
$n$ argument terms $f(t_1, \ldots, t_n)$. 

A \emph{rewrite rule} is a pair of terms
$\ell \to r$ and a TRS~$\RR$ is a set of rewrite rules.
The \emph{rewrite relation (induced by $\RR$)} $\rstep$ is the closure
under substitutions and under contexts of $\RR$, i.e., $s \rstep t$ iff
there is a context~$C$, a rewrite rule $\ell\to r \in \RR$, and a substitution
$\sigma$ such that
$s = C[\ell\sigma]$ and $t = C[r\sigma]$. 
A TRS~$\RR$ is terminating, written $\SN(\RR)$, if there is no
infinite derivation $t_1 \to_\RR t_2 \to_\RR t_3 \to_\RR \dots$.

\newcommand{\pl}{\oplus}
\newcommand{\ti}{\odot}
\newcommand{\one}{\underline{\mathsf{1}}}
\newcommand{\zero}{\underline{\mathsf{0}}}
\newcommand{\mono}{\mathsf{mono}}
\newcommand{\pos}{\mathsf{pos}}
\renewcommand{\max}{\mathsf{max}}
\newcommand{\mat}[1]{#1^{n \times n}}
\newcommand{\matsd}[1]{#1_{sd}^{n \times n}}

\section{Polynomial Orders with Negative Constants}
\label{snp}
Polynomial orders \cite{L79} are a well-known technique to prove
termination. They are an instance of the termination technique of
well-founded monotone algebras. Such algebras can be used for all
termination techniques that rely on \emph{reduction pairs} \cite{AG00}.
Here, a reduction pair consists of two partial orders $(\succsim,\succ)$
where $\succsim$ and $\succ$ are stable, $\succsim$ is reflexive and
monotone, $\succ$ is well-founded, and $\succsim$ is compatible to $\succ$,
i.e., ${\succsim} \circ {\succ} \subseteq {\succ}$. If additionally $\succ$
is monotone, then we call $(\succsim,\succ)$ a \emph{monotone} reduction
pair.

It is well-known that reduction pairs can be used for proving termination
of TRSs within the DP framework \cite{AG00,JAR06,HM05}.
Moreover, monotone reduction pairs can be used for direct termination
proofs or rule removal \cite{CL87,Geser90,L79}.

To formalize polynomial orders, we first assume some semiring over which the 
polynomials are built. 
\begin{definition}
\label{def:semiring1}
A structure $(\UU,\pl,\ti,\zero,\one)$ with universe $\UU$, two binary
operation $\pl$ and $\ti$ on $\UU$, and with $\zero, \one \in \UU$ is a
\emph{semiring with one-element} iff
\begin{itemize}
\item $\pl$ and $\ti$ are associative and $\pl$ is commutative
\item $\zero \neq \one$, $\zero$ and $\one$ are neutral elements w.r.t.~$\pl$ and $\ti$,
  respectively, and $\zero \ti x = x \ti \zero = \zero$
\item $\ti$ distributes over $\pl$: $x \ti (y \pl z) = x \ti y \pl x \ti z$
and $(x \pl y) \ti z = x \ti z \pl y \ti z$ 
\end{itemize}
\end{definition}
To obtain polynomial orders, we assume a strict and a non-strict order.
Moreover, we demand the existence of a unary predicate $\mono$ where
$\mono(x)$ indicates that multiplication with $x$ is monotone w.r.t.~the
strict order.

\begin{definition}
A structure $(\UU,\pl,\ti,\zero,\one,\geq,>,\mono)$ is an \emph{ordered
semiring} iff $(\UU,\pl,\ti,\zero,\one)$ is a semiring with one-element and
additionally:
\begin{itemize}
\item $\geq$ is reflexive and transitive;
$>$ and $\geq$ are compatible: ${>} \circ {\geq} \subseteq {>}$ and ${\geq} \circ {>} \subseteq {>}$
\item $\one \geq \zero$ and $\mono(\one)$
\item $\pl$ is left-monotone w.r.t.~$\geq$: if $x \geq y$ then $x \pl z \geq y
\pl z$
\item 
$\pl$ is left-monotone w.r.t.~$>$: if $x > y$ then $x \pl z > y \pl z$
\item $\ti$ is left-monotone w.r.t.~$\geq$: if $x \geq y$ and $z \geq \zero$
then $x \ti z \geq y \ti z$; $\ti$ is right-monotone w.r.t.~$\geq$
\item $\ti$ is right-monotone w.r.t.~$>$: if $\mono(x)$, $x \geq \zero$, and $y > z$
then $x \ti y > x \ti z$
\item $\{(x,y) \mid x > y \wedge y \geq \zero\}$ is well-founded
\end{itemize}
\end{definition}

Note that using the approach of well-founded monotone algebras, every
\pagebreak
interpretation of the function symbols over some ordered semiring gives
rise to a strict ($\succ$) and a non-strict ($\succsim$) order on terms.
For example, for a 
polynomial interpretation $\Pol$ we define $s
\succ_{\Pol} t$ iff $[s] > [t]$, and $s \succsim_{\Pol} t$ iff $[s] \geq
[t]$ where $[s]$ is the homeomorphic extension of $\Pol$ to terms. 

\begin{theorem}
\label{polyorder}
Let $\Pol$ be a polynomial interpretation over an ordered semiring
$(\UU,\pl,\ti,\zero,\one,\geq,>,\mono)$ where $[f](x_1,\dots,x_n) = f_0 \pl
f_1 \ti x_1 \pl \cdots \pl f_n \ti x_n$ and $f_i \geq \zero$ for all $0 \leq
i \leq n$ and every $n$-ary symbol $f$.  Then
$(\succsim_{\Pol},\succ_{\Pol})$ is a reduction pair.  If moreover,
$\mono(f_i)$ for all $1 \leq i \leq n$ then
$(\succsim_{\Pol},\succ_{\Pol})$ is a monotone reduction pair.
\end{theorem}

\begin{example}[Ordered Semirings]
\label{basic domains}
$(\nats, {+}, {\cdot}, 0, 1, {\geq}, {>}, {\geq 1})$,
$(\ints, {+}, {\cdot}, 0, 1, {\geq}, {>}, {\geq 1})$, and 
$(\rats, {+}, {\cdot}, 0, 1, {\geq}, \linebreak{>_\delta}, {\geq 1})$
are ordered semirings. In the last case, we assume a fixed rational number
$\delta$ with $0 < \delta$, and where $>_\delta$ is defined by $x
>_\delta y$ iff $x - y \geq \delta$.
\end{example}

To formalize matrix-interpretations \cite{MatrixJAR}, we followed the
approach of \cite{MatrixCoq} and used a domain with an additional
strict-dimension and where the elements are matrices---instead of vectors
as in \cite{MatrixJAR}. In detail, we have proven that if $0 < sd \leq n$
and
$(\UU,\pl,\ti,\zero,\one,{\geq},{>},\mono)$ is an ordered semiring, then
$(\mat\UU,\mat\pl,\mat\ti,\mat\zero,\mat\one,\mat\ge,\matsd>,\matsd\mono)$ 
is also an ordered semiring where all operations and constants are lifted
to work on $n$-dimensional matrices, with the strict-dimension $sd$.  Here,
$\mat\ge$ compares the arguments component-wise, and $M \matsd> M'$ iff $M
\mat\ge M'$ and at least one entry in the upper-left $sd \times
sd$-submatrix is strictly decreasing w.r.t.~$>$.  Moreover, $\matsd\mono$
demands that for every column in the upper-left $sd \times sd$-submatrix
there is at least one monotone entry. 

As observed in \cite{MatrixCoq}, choosing $sd = 1$, is comparable to the
classic definition of matrix-interpretations. Choosing $sd = n$, is always
best if one does not require monotonic reduction pairs. However, to ensure
monotonicity also a small value of $sd$ might be attractive. 

\newcommand{\pleft}[1]{[#1]_{\mathit{left}}}
\newcommand{\pright}[1]{[#1]_{\mathit{right}}}
\newcommand{\cp}{\mathsf{cp}}
\newcommand{\ncp}{\mathsf{ncp}}

To lift the requirement in \thmref{polyorder} that all $f_i$ have to be at
least $\zero$, in \cite{HM07}, polynomial orders with negative constants
have been introduced. There, the constant part can be arbitrary but the
interpretation of a function is always wrapped into a $\max(\zero,\cdot)$
operation to ensure well-foundedness. This complicates the comparison of
terms, as the resulting interpretations are not pure polynomials anymore,
but also contain the $\max$-operator. To this end, approximations
$\pleft\cdot$ and $\pright\cdot$ have been introduced which interpret terms
by polynomials without $\max$, such that $\pleft{s} \leq [s] \leq
\pright{s}$.

However, the existing approximations are unsound if generalized naively.
For example, in the case where the constant part is negative, it is
removed. This works fine for the integers and the rationals, but not for
matrices, as here some parts of the matrix may be negative, but other parts
can also be positive and thus, cannot be removed. Thus, we formalized the
following approximations which are equivalent to those of \cite{HM07}, but
also work for matrices:

\begin{definition}
\label{leftright} Let $\cp(\cdot)$ be the constant part and $\ncp(\cdot)$ be
 the non-constant part of a polynomial.
\begin{align*}
\pleft{x} = \pright{x}  & = x \\
\pleft{f(t_1,\dots,t_n)} & = \begin{cases}
\max(\zero,\cp(p_{\mathit{left}}))
  & \text{if $\ncp(p_{\mathit{left}}) = \zero$} \\
p_{\mathit{left}} & \text{otherwise}
\end{cases} \\
\pright{f(t_1,\dots,t_n)} & =  \ncp(p_{\mathit{right}}) \pl \max(\zero,\cp(p_{\mathit{right}}))\\
\text{where } p_{\mathit{left}} & = [f](\pleft{t_1},\dots,\pleft{t_n}) \quad
\text{and }\quad p_{\mathit{right}} = [f](\pright{t_1},\dots,\pright{t_n}) 
\end{align*}
\end{definition}

Note that for \defref{leftright} we have to extend ordered semirings by the
additional unary operation: $\max(\zero,\cdot)$. 

\newcommand{\maxz}{\mathsf{max}\zero}

\begin{definition}
A structure $(\UU,\pl,\ti,\zero,\one,\geq,>,\mono,\maxz)$ is an
\emph{ordered semiring with max} iff
$(\UU,\pl,\ti,\linebreak\zero,\one,\geq,>,\mono)$ is an ordered semiring
and additionally:
\begin{itemize}
\item $\maxz(x) \geq \zero$ and $\maxz(x) \geq x$
\item $y \geq x \geq \zero$ implies $\maxz(y) \geq \maxz(x) = x$\pagebreak
\end{itemize}
\end{definition}

\begin{theorem}
\label{negpolyorder}
Let $(\UU,\pl,\ti,\zero,\one,\geq,>,\mono,\maxz)$ be an ordered semiring
with max and $\Pol$ be a polynomial interpretation where
$[f](x_1,\dots,x_n) = f_0 \pl f_1 \ti x_1 \pl \cdots \pl f_n \ti x_n$ and
$f_i \geq \zero$ for all $1 \leq i \leq n$ and every $n$-ary symbol $f$.
Then $(\succsim_{\Pol},\succ_{\Pol})$ is a reduction pair where $s \succ /
\succsim t$ can be approximated by $\pleft s > / \geq \pright t$.
\end{theorem}

\begin{example}
All ordered semirings of \exref{basic domains} are also ordered semirings
with max, where $\maxz$ is the standard operation on $\nats$, $\ints$, and
$\rats$, and $\maxz$ is performed component-wise for matrices.

For example, for $\rats$ it is now possible to use interpretations like
\begin{align*}
[\Fhalf](x) & = \frac12 \cdot x + \frac12 & 
[\Fp](x) & = x - 1 &
[\Fs](x) & = x + 1
\end{align*}
where 
\[
\pleft{\Fs(x)} = x + 1 >  \frac12 
\cdot x + \frac12 = \pright{\Fp(\Fhalf(\Fs(\Fs(x))))}
\]
Since we are not aware of any termination tool that supports these interpretation,
we would like to encourage their integration, perhaps an interpretation like
\[
[\Ff](x,y) = \matm{\frac12}8{\frac75}0 \cdot x 
     + y + \matm{-\frac13}3{-5}{\frac29}
\]
increases the power in the next competition.

\end{example}

\section{Arctic Semirings}
\label{sarc}
In \cite{Arctic}, the \emph{arctic semiring} as well as the \emph{arctic
semiring below zero}, where used the first time in the well-founded
monotone algebra setting.

\begin{example}[Arctic Semirings]
\label{ex:arcsemi}
The arctic semiring $(\arcnats,\max,+,\arczero,\arcone)$, the arctic
semiring below zero $(\arcints,\max,+,\arczero,\arcone)$, and the arctic
rational semiring $(\arcrats,\max,+,\arczero,\arcone)$, are semirings with
one-element as in \defref{def:semiring1}. The carriers are given by
$\mathbb{A}_S = S \cup \{\arczero\}$.  Furthermore, the standard operations
$\max$ and $+$ are extended such that $\max\{x,\arczero\} = x$ and $x +
\arczero = \arczero + y = \arczero$, for all $x$ and $y$.
\end{example}

\begin{definition}
A structure $(\UU,\pl,\ti,\zero,\one,\geq,>,\pos)$ is an ordered arctic
semiring iff $(\UU,\pl,\ti,\zero,\one)$ is a semiring with one-element and
additionally:
\begin{itemize}
\item $\geq$ is reflexive and transitive; $>$ and $\geq$ are compatible:
  ${>} \circ {\geq} \subseteq {>}$ and ${\geq} \circ {>} \subseteq {>}$
\item $\one \geq \zero$; $\pos(\one)$; $x > \zero$; $x \geq \zero$;
  and if $\zero > x$ then $x = \zero$
\item $\pl$ is left-monotone w.r.t.\ $\geq$
\item $\pl$ is monotone w.r.t.\ $>$: if $x > y$ and $x' > y'$ then $x \pl x' > y \pl y'$
\item $\ti$ is left- and right-monotone w.r.t.\ $\geq$ and left-monotone
w.r.t.\ $>$
\item staying positive: if $\pos(x)$ and $\pos(y)$ then $\pos(x \pl z)$ and
  $\pos(x \ti y)$
\item $\{(x, y) \mid x > y \land \pos(y)\}$ is well-founded
\end{itemize}
\end{definition}

\begin{theorem}
\label{arcpoly}
Let $\Pol$ be a polynomial interpretation over an ordered arctic semiring
$(\UU,\pl,\ti,\zero,\one,\geq,\linebreak>,\pos)$ where $[f](x_1,\ldots,x_n)
= f_0 \pl f_1 \ti x_1 \pl \cdots \pl f_n \ti x_n$ and $\pos(f_i)$ for some
$0 \leq i \leq n$ and every $n$-ary symbol $f$. Then
$(\succsim_{\Pol},\succ_{\Pol})$ is a reduction pair where $s \succsim /
\succ t$ is approximated by comparing $[s]$ and $[t]$ component-wise using $> / \geq$.
 (For example to compare $a \ti x \pl b \ti y \pl c > d \ti x \pl e \ti y \pl f$ one demands $a > d$,  $b > e$, and $c > f$.)
\end{theorem}

Moreover, if $n > 0$ and $(\UU,\pl,\ti,\zero,\one,\geq,>,\pos)$ is an
ordered arctic semiring, then
$(\mat\UU,\mat\pl,\linebreak\mat\ti,\mat\zero,\mat\one,\mat\geq,\mat>,\mat\pos)$
is also an ordered arctic semiring where all operations and constants are
lifted to work on $n$-dimensional matrices. Here $\mat\geq$ and $\mat>$,
compare arguments component-wise and $\mat\pos$ checks, whether the
leftmost topmost element is $\pos$.

\begin{example}[Ordered Arctic Semirings]
All arctic semirings of \exref{ex:arcsemi} are also ordered arctic
semirings. In all three cases, we use the non-strict ordering $x \geq y
\equiv y = \arczero \lor (x \neq \arczero \land x \geq_{\nats/\ints/\rats} y)$.  For
$\arcnats$ and $\arcints$, we use the strict ordering $x > y \equiv y =
\arczero \lor (x \neq \arczero \land x >_{\nats/\ints} y)$, and for $\arcrats$,
we use the strict ordering $x >_\delta y \equiv y = \arczero \lor (x \neq
\arczero \land x - y \geq_{\rats} \delta)$ for some $\delta > 0$. 
Furthermore, the check for
positiveness is defined by $\pos(x) \equiv x \neq \arczero \land x
\geq_{\nats/\ints/\rats} 0$.
\end{example}

Note that the ordered arctic semiring over $\arcrats$, together with
\thmref{arcpoly}, unifies and extends Theorems~12 and~14 of \cite{Arctic}.
Here, the main advantage of our approach is that we only restrict interpretations 
$[f](x_1,\dots,x_n) = f_0 \pl f_1 \ti x_1 \pl \cdots \pl f_n \ti x_n$ by demanding
that at least one $f_i$ is positive. This is in contrast to the theorem
about the arctic semiring below zero in \cite{Arctic} 
where always the constant part $f_0$ has to
be positive. However, Waldmann observed that for finite TRSs one can 
transform every polynomial order over the arctic rationals into an order over the 
arctic naturals by multiplication and shifting.


%
\bibliographystyle{plain}
\bibliography{wst10}

\end{document}